# The stationary state and gravitational temperature in a pure self-gravitating system

Zheng Yahui[1,2], Du Jiulin[1]

1: *Department of Physics, School of Science, Tianjin University, Tianjin 300072, China*
2: *Department of Physics, School of Science, Qiqihar University, Qiqihar City 161006, China.*

**Abstract** The pure self-gravitating system in this paper refers to a multi-body gaseous system where the self-gravity plays a dominant role and the intermolecular interactions can be neglected. Therefore its total mass must be much more than a limit mass, the minimum mass of the system exhibiting long-range nature. Thee method to estimate the limit mass is then proposed. The nonequlibrium stationary state in the system is identical to the Tsallis equilibrium state, at which the Tsallis entropy approaches to its maximum. On basis of this idea, we introduce a new concept of the temperature whose expression includes the gravitational potential and therefore we call it gravitational temperature. Accordingly, the gravitational thermal capacity is also introduced and it can be used to verify the thermodynamic stability of the astrophysical systems.

**1. Introduction**

The self-gravitating system is a type of multi-body system naturally organized through the gravitational interactions. Planets like our Earth, stars like the Sun, and even the galaxies consisting of so many stars all belong to such systems. It is well known that in many cases the classical Boltzmann-Gibbs (BG) statistical method is not applicable to such astrophysical systems because the exponential distribution functions based on BG statistics would lead to many unreasonable theoretical results, such as the infinite mass and the negative capacity problem.

Since nonextensive statistics was proposed in 1988 [1], the theoretical successes in both astronomy & astrophysics [2-5] and space plasmas [6-11] have shown that the new statistical method may be suitable for describing many astrophysical systems. In this work, we apply the methods of nonextensive statistics to study the nonequilibrium property of a pure self-gravitating system. The pure self-gravitating system refers to a gaseous system in which no phase transition and no nuclear reactions take place and the self-gravity plays a dominant role.

Obviously, such a pure self-gravitating system is a gaseous sphere consisting of the particles with self-gravitating interactions. In the classical BG statistics, the total energy of the gaseous sphere [12] is

$$E = -\frac{3}{2} N k_B \overline{T},\qquad(1)$$

and so the thermal capacity is

$$C_V = \frac{dE}{d\overline{T}} = -\frac{3}{2} N k_B,\qquad(2)$$

where $\overline{T}$ is the average temperature of the system. Eq.(2) indicates that the system is thermodynamically unstable and will lead to the so-called gravothermal catastrophe, which was discussed in the system in a spherical adiabatic wall [13,14]. According to Eq.(2), the conclusion can be drawn that most self-gravitating systems do not exist in the universe for long-term, which, obviously, conflicts with the observation facts.

In the previous work, we divided the self-gravitating system as three kinds. In the first kind of system whose total mass is less than $\sigma M_e$, where $M_e$ is the mass of the earth and $\sigma$ is a parameter whose value ranges about 10~300 [15], the intermolecular potential can plays an important role. Under certain conditions, say, when the pressure is high enough near the center of the system, the phase transition related to the intermolecular interaction may take place, one result of which is the sign of the thermal capacity of the whole system to become positive, thus leading to thermodynamic stability of this kind of system. In the third kind of system whose total mass is more than $0.08 M_\odot$, where $M_\odot$ is the solar mass, just as everyone knows, there exist nuclear reactions in the core. Obviously, the intermolecular interactions can be ignored due to very high temperature in such case. Therefore, the phase transition similar to that in the first kind of system does not appear. Although its heat capacity is negative, the energy compensation due to the nuclear reactions keeps thermodynamic stability of the whole system, which actually presents almost in all of the stars. In the second kind of system whose total mass is more than $\sigma M_e$ and less than $0.08 M_\odot$, the short-range intermolecular potentials can also be neglected. It is clear that in such a system there is neither the phase transition related to the molecular potentials nor the nuclear reactions taking place. This kind of system is just the pure self-gravitating system we discussed in this paper.

The paper is organized as follows. The minimum mass as the pure self-gravitating system is studied in section 2, the nonequilibrium stationary state and the Tsallis equilibrium is discussed in section 3, the concept of gravitational temperature is introduced in section 4, the thermodynamic stability condition (TSC) and the gravitational thermal capacity are studied in section 5, and finally the conclusions and discussions are give in section 6.

## 2. The minimum mass as the pure self-gravitating system

It is well-known that the intermolecular potentials have the nature of short-range interactions, then in the system consisting of many molecules, the total intermolecular potential is approximately proportional to the molecule number $N$. In contrast, the gravitational interactions have the nature of long-range, then the total gravitational potential is approximately proportional to some power of the molecule number $N$ of the system, that is, $\sim N^n$, where n is a real number and n>5/3 [15] (This is a conservative estimate we give according to the constant density distribution.

For a self-gravitating system with constant density, its total potential energy (absolute value) should be proportional to $N^{5/3}$. For a real self-gravitating system whose mass is more concentrated, the total potential energy is more than that of the system with constant density. This means the exponent may be n>5/3.)

Obviously, with the increase of molecule number, the total gravitational potential increases more rapidly than the total molecular potential in the gaseous system. When the molecule number is large enough, the intermolecular potentials can be neglected compared with the gravitational potentials, thus the latter will play a dominant role in such system. Therefore, we now propose a method to estimate the minimum mass of such a pure self-gravitating system consisting of gaseous molecules.

For the laboratory system whose size is about dozens of meters, the self-gravitating potential energy can be ignored relatively to the interaction energy between the molecules. On the contrary, for the huge self-gravitating gaseous system, like the Sun, the molecular potentials are negligible relatively to the gravitational potential energy. Therefore, the minimum mass can be determined by just equating these two potential energies. For simplicity, we assume that the mass density of the pure self-gravitating system is a constant, and then we easy find the self-gravitating potential energy,

$$U = -\frac{6}{5}\frac{GM^2}{R}, \qquad (3)$$

where $M$ is the total mass of the system, and $R$ is the radius of the gaseous sphere. Next, for calculating the interaction energy of molecules, we adopt the two-molecule interaction potential proposed by Lenard Jones [16],

$$\phi(r) = \phi_0[(\frac{r_0}{r})^{12} - 2(\frac{r_0}{r})^6], \qquad (4)$$

where $r$ is the distance between any two molecules, $\phi_0$ and $r_0$ are the characteristic parameters determined by the experiments. Supposing that the molecules are at their equilibrium positions, $r=r_0$, the total interaction potential energy of the molecules is $-N\phi_0/2$. Equating these two potential energies, one has

$$\frac{6}{5}\frac{GM^2}{R} = \frac{N\phi_0}{2}. \qquad (5)$$

The volume of the system can be written as

$$\frac{4}{3}\pi R^3 = N\frac{4}{3}\pi r_0^3. \qquad (6)$$

Setting $M=Nm$, and combining Eq.(5) with Eq.(6), one can find the minimum mass,

$$M_L = \frac{1}{m^2}(\frac{5r_0\phi_0}{12G})^{3/2}, \qquad (7)$$

where $m$ is the molecule mass. This formula (7) gives the minimum mass of the pure self-gravitating system. To see this point more clearly, we take the argon gaseous sphere as an example. For this sphere, one has $r_0 = 3.82\times 10^{-10} m$, and $\phi_0/k_B$=120K [16]. Then the minimum mass of the pure self-gravitating argon sphere is $M_L = 5.7\times 10^{19} kg$. This value is only about $1/10^5$ of the Earth's mass.

If the gaseous system has such a limit mass, the molecular potential and the gravitational potential are equal to each other. If its total mass is less than the limit mass, the total energy of the system tends to be positive, so it is not a self-gravitating system according to the definition. Obviously, this limit mass $M_L$ is the minimum mass of the first kind of self-gravitating system. As the pure self-gravitating system, where the short-range potential can be ignored entirely, its mass must be much more than this limit mass. At least, it should have the order of $10^7 M_L$ (about 100 times the earth's mass, which was derived from the reference range of σ in [15]), which is the upper limit of the mass of the first kind of the system. Beyond this value, no matter how high the pressure in the system center is, the molecular potentials are always unable to arouse the phase transition process. Just in this sense we say that the pure self-gravitational system discussed in this paper is of the second kind of self-gravitating system [15].

## 3. The nonequilibrium stationary state and Tsallis equilibrium

Now we discuss the equivalence between the nonequilibrium stationary state and the Tsallis equilibirum in a pure self-gravitating system. The pure self-gravitating system, if there is not any heat source around it apart from the vacuum, can be regarded as a isolated system. For such a self-gravitating system, the total Tsallis entropy can be written as

$$S_q(f) = -k \int f^q \ln_q f d^3\vec{v} d^3\vec{r} . \tag{8}$$

The distribution function $f(\vec{v}, \vec{r})$ in this equation satisfies the generalized Boltzmann equation in the nonextensive statistical framework [17],

$$\frac{\partial f}{\partial t} + \vec{v} \cdot \frac{\partial f}{\partial \vec{r}} + \frac{\vec{F}}{m} \cdot \frac{\partial f}{\partial \vec{v}} = C_q(f) , \tag{9}$$

which is of course appropriate for this self-gravitating system, where $C_q$ is the $q$-collision term. This equation can describe a self-gravitating gaseous sphere of $N$ hard-sphere particles of mass $m$ and diameter $s$. Here the force $\vec{F}$ is not the external force but the self-gravity in the system, so one has $\vec{F}/m = -\nabla\varphi$, where $\varphi$ is the gravitational potential. The $q$-collision term in Eq.(9) can be expressed as

$$C_q(f) = \frac{s^2}{2} \int |\vec{V} \cdot \vec{e}| R_q d\Omega d^3\vec{v}_1 , \tag{10}$$

where $d^3\vec{v}_1$ stands for the volume element in velocity space, $\vec{V}$ is the relative velocity before collision, $\vec{e}$ is an arbitrary unit vector, and $s^2 d\Omega$ is the area of collision cross section. $R_q$ is the difference of correlated distribution functions, which are assumed to satisfy the generalized form of molecular chaos hypothesis [17],

$$R_q(f, f') = e_q(f'^{(q-1)} \ln_q f' + f_1'^{(q-1)} \ln_q f_1') - e_q(f^{(q-1)} \ln_q f + f_1^{(q-1)} \ln_q f_1) , \tag{11}$$

where the $q$-logarithmic function is defined as

$$\ln_q f = \frac{f^{1-q} - 1}{1 - q} \tag{12}$$

and the $q$-exponential function is

$$e_q(f) = [1 + (1-q)f]^{\frac{1}{1-q}} . \tag{13}$$

The time derivative of Tsallis entropy is

$$\frac{dS_q}{dt} = -k\int (1+qf^{q-1}\ln_q f)\frac{\partial f}{\partial t}d^3\vec{v}d^3\vec{r}. \tag{14}$$

Substituting Eq.(9) into the above equation, one obtains

$$\frac{dS_q}{dt} = -k\int (1+qf^{q-1}\ln_q f)C_q(f)d^3\vec{v}d^3\vec{r} + k\int \nabla\cdot(\vec{v}f^q\ln_q f)d^3\vec{v}d^3\vec{r}$$
$$+k\int \nabla_{\vec{v}}\cdot(\frac{\vec{F}}{m}f^q\ln_q f)d^3\vec{v}d^3\vec{r}. \tag{15}$$

On the right side of Eq.(15), the second and third terms can be changed into the area integral in the position space and velocity space, respectively. The distribution function tends to zero at the boundary of such spaces. Therefore, Eq.(15) becomes

$$\frac{dS_q}{dt} = -k\frac{s^2}{2}\int (1+qf^{q-1}\ln_q f)|\vec{V}\cdot\vec{e}|R_q d\Omega d^3\vec{v}_1 d^3\vec{v}d^3\vec{r}. \tag{16}$$

According to the collision symmetry and along the standard step [17], Eq.(16) can be further modified as

$$\frac{dS_q}{dt} = k\frac{s^2 q}{8}\int (f'^{(q-1)}\ln_q f' + f_1'^{(q-1)}\ln_q f_1' - f^{(q-1)}\ln_q f - f_1^{(q-1)}\ln_q f_1)$$
$$\times |\vec{V}\cdot\vec{e}|R_q d\Omega d^3\vec{v}_1 d^3\vec{v}d^3\vec{r}. \tag{17}$$

Considering Eq.(11), one can find immediately the result that

$$\frac{dS_q}{dt} \geq 0, \quad \text{for q>0,} \tag{18}$$

if and only if the following equation holds,

$$f'^{(q-1)}\ln_q f' + f_1'^{(q-1)}\ln_q f_1' = f^{(q-1)}\ln_q f + f_1^{(q-1)}\ln_q f_1. \tag{19}$$

Eq.(18) is just the second law of thermodynamics. When the equal sign holds in Eq.(18), the Tsallis entropy approaches its maximum value and the whole system will be at the Tsallis 'equilibrium' state, where one can formulate the single-molecule distribution function for describing the equilibrium property of the isolated self-gravitating system.

According to the second law of thermodynamics, Eq.(18), when the condition (19) is satisfied, the pure self-gravitating system, which is isolated, should be at the state with the maximum value of Tsallis entropy. The maximum value state of the entropy is ordinarily called Tsallis equilibrium state. Furthermore, it is believed that such isolated self-gravitating systems, when they are not disturbed and the convective instability does not take place inside them, will ordinarily be at the hydrostatic equilibrium, satisfying the hydrostatic equilibrium equation,

$$\nabla P = -mn\nabla\varphi \tag{20}$$

where $P$ is the pressure inside the self-gravitating gaseous sphere. According to the viewpoint of thermodynamics, the hydrostatic equilibrium is a nonequilibrium stationary state. The Tsallis equilibrium distribution for the self-gravitating system describes the property of the system at the hydrostatic equilibrium [18].

In the above discussion, we neglect the energy loss of the system due to the thermal radiation.

The pure self-gravitating system with no heat source surrounded can be regarded as isolated one. For such one isolated self-gravitating system, if it is at the hydrostatic equilibrium, which is also the nonequilibrium stationary state, it must have the definite energy, velocity distribution function and temperature distribution. Furthermore, now that the gravity is the internal force, the self-gravitating potential can stop the heat conduction related to the temperature gradient inside the system, so under certain condition the total entropy can be at its maximum value. In this sense we can say the nonequilibrium stationary state is identical to the Tsallis equilibrium state where the Tsallis entropy approaches its maximum value. We would see in next section, this equivalence property is very important to our discussions.

**4. The gravitational temperature**

For an isolated pure self-gravitating system, the Tsallis equilibrium state is equivalent to the nonequilibrium stationary state. This means that there must be one state parameter, which has fixed value in the whole system and can mark the state (Tsallis equilibrium) of such a system, with the definite total energy and the Tsallis entropy. Of course, the best choice for this state parameter is temperature. However, it is well-known that the temperature in self-gravitating gas systems is inhomogeneous. Therefore, how it can be used to describe the stationary state?

One remedy method may be to employ the average value of the inhomogeneous temperature. But the average value will lose the distribution information about the inhomogeneous nature, and is unable to determine or mark the state of this system. Fortunately, we have confirmed that, for the nonextensive system like the self-gravitating one, the concept of temperature could generally split into two: the physical temperature related to the nonextensivity of the whole system [19], and the Lagrange temperature (the inverse of Lagrange multiplier) related to the local molecular collisions. The former is used to describe the 'equilibrium' state of the nonextensive system, and the latter is the measurable quantity in experiments. Their roles are different, but both useful and indispensable. This is just the assumption of temperature duality [20]. In a pure self-gravitating system, we also employ this temperature assumption, and similar to the definition of physical temperature in the ensemble theory [19] one can introduce a new concept of the generalized temperature for the molecular dynamics.

In order to carry out this work, we must point out a basic property of the physical temperature in an isolated system. As the state parameter to describe the 'equilibrium' state, it must be homogeneous everywhere. Setting it with $T_g$, this means

$$\nabla T_g = 0 \tag{21}$$

Furthermore, as the whole variable it should be related to the nonextensivity of the whole system. In the case of self-gravitating system, the nonextensivity behaves mainly through the self-gravity, therefore this quantity should be related to the gravitation of the system. Actually, just in this sense we can call the parameter as the gravitational temperature.

Now let us introduce the Tsallis equilibrium distribution, or the generalized Maxwellian velocity distribution, into the isolated pure self-gravitating system at the nonequilibrium stationary state. According to Eq.(19), let $q-1=1-q^*$, the distribution function at Tsallis equilibrium can be expressed [21] as

$$f_{q*}(\vec{r},\vec{v}) = nB_{q*}(\frac{m}{2\pi kT})^{\frac{3}{2}}[1-(1-q*)\frac{mv^2}{2kT}]^{\frac{1}{1-q*}}, \text{ for } q>0, \qquad (22)$$

where $n$ is the number density, $m$ is the particle's mass, $T$ is the Lagrange temperature, and $B_{q*}$ is the normalized constant related to the nonextensive parameter $q$. For convenience, let $Q=1-q*$, then the above $q$-distribution can be written as

$$f_Q(\vec{r},\vec{v}) = nB_Q(\frac{m}{2\pi kT})^{\frac{3}{2}}[1-Q\frac{mv^2}{2kT}]^{\frac{1}{Q}}. \qquad (23)$$

One of the important conclusions about the Tsallis $q$-distribution is the relationship derived for the nonextensive parameter $q$ [21],

$$k\nabla T + Qm\nabla \varphi = 0 \qquad (24)$$

with the condition

$$\nabla Q = 0 \qquad (25)$$

In view of Eqs. (24) and (25), we can introduce the gravitational temperature $T_g$,

$$T_g = T + \frac{Qm\varphi}{k} + T_0, \qquad (26)$$

where the constant $T_0$ is determined by the 'equilibrium' condition. One can find the condition (21) is satisfied naturally at the 'equilibrium' state. This expression is different from the physical temperature defined from the ensemble theory [19,20]. The gravitational temperature defined here is the molecular kinetics form of the physical temperature in the pure self-gravitational system and therefore is identical to physical temperature. Obviously, the definition in Eq.(26) is superposition of the Lagrange temperature $T$ and the temperature associated with the gravitational potential. It is closely related to the assumption of temperature duality.

Generally, the dominant term in Eq.(26) is the Lagrange temperature $T$. For example, in the laboratory system, the gravitational potential as an external field is thought a constant (along the horizontal direction). According to Eq.(26), this means that when the system is at thermal equilibrium, the gravitational temperature and the Lagrange temperature are the same everywhere. However, in the pure self-gravitational system, the gravitational potential, as the effective potential in the mean field theory, depends on position and thus is different everywhere in the system. Therefore, the gravitational temperature is not the same as the Lagrange temperature. The gravitational temperature is a generalized form of the Lagrange temperature (the thermodynamic temperature) when the potential is not a constant in the system.

From Eq.(26) we know that at the Tsallis equilibrium the gravitational temperature is space homogeneous everywhere. So $T_g$ is the state parameter describing the 'equilibrium' property of the pure self-gravitating system, and it is responsible for the "thermal" balance between the different parts of the closed system. Of course, whether the balance is stable is determined by the sign of the specific heat of the self-gravitating system. This will be discussed in the next section.

Notice that the gravitational potential in Eq.(26) is negative, now that the zero potential point is infinite far away. So it is easy to see that, for the isolated system, the inhomogeneous distribution of the Lagrange temperature $T$ can be kept for long time, no matter how high the temperature is in the center. Of course, here we neglect the effect of the thermal radiation, which is determined by the Lagrange temperature *T*. Actually, the radiation energy exchange of the system with the surroundings leads to slow energy loss of the self-gravitating system. This is why the celestial body as one quasi-isolated system can evolve for long time. In the next section, we will use the gravitational temperature to show the stability of the 'equilibrium' state of the pure self-gravitating system.

**5. The TSC and the gravitational thermal capacity**

According to Eq.(1) and Eq.(2), the negative thermal capacity can be derived from the definition with the total energy of the system, i.e. the internal energy (ordinarily understood as the sum of the kinetic energies of all molecules) and the total gravitational potential energy. The virial theorem [22] shows that, due to the long-range nature, the gravitational potential is dominant in the total energy. Therefore, when the total energy of the pure self-gravitating system is low sufficiently, its thermal capacity is negative. Just as we mentioned above, the negative thermal capacity does not imply instability of the system. We need to find another thermal capacity which relates the gravitational temperature and determines the stability of the system. The obvious and feasible method to construct such a quantity may be to replace the Lagrange temperature differential element by the gravitational temperature differential element. For this purpose, let us consider the TSC of Tsallis entropy.

Generally speaking, the TSC of Tsallis entropy can be written [23] as

$$\frac{k+(1-q)S_q}{C_V}+1-q \geq 0, \text{ for } q>0, \tag{27}$$

For the self-gravitating system, the thermal capacity $C_V$ in Eq.(27) should be the same as Eq.(2), and therefore it is negative. Furthermore, the relation between the gravitational temperature (i.e. the physical temperature in nonextensive statistics [19,20]) and the Lagrange temperature is

$$T_g = [1+(1-q)S_q/k]T. \tag{28}$$

Similar to the equation (2), we can introduce the gravitational thermal capacity $C_{Vg}$, defined as

$$C_{Vg} = \frac{dE}{dT_g}. \tag{29}$$

Based on Eq.(28), the TSC (27) can be written [23] as

$$\frac{1}{C_{Vg}} = \frac{k+(1-q)S_q}{C_V}+1-q \geq 0. \tag{30}$$

This inequality implies that the TSC of the Tsallis equilibrium state can be expressed only by the gravitational thermal capacity. Apparently, when this new thermal capacity is positive, the self-gravitating system is stable, but when it is negative, the system is unstable. Therefore, the gravitational thermal capacity can determines the stability of the self-gravitating system. This is consistent with the analysis in the former section: The new thermal capacity depends on the gravitational temperature only, which determines the property of the Tsallis equilibrium.

Now we discuss the property of the gravitational thermal capacity in the molecular kinetics of the pure self-gravitating system. The average of Eq.(26) for the whole system is

$$NkT_g = Nk\overline{T} + QN\overline{m\varphi} + NkT_0. \tag{31}$$

According to the virial theorem [22], one has

$$\sum_i m_i \varphi_i = N \overline{m\varphi} = -2U \tag{32}$$

Here, as one mean filed, the total gravitational potential of the system is the sum of the potential energies of single molecules. The quantity $U$ is the total molecular energy of the self-gravitating system. If the freedom of the molecule is $D$, one has

$$U = \frac{D}{2} N k \overline{T}, \tag{33}$$

and then,

$$T_g = (1 - DQ)\overline{T} + T_0. \tag{34}$$

It is easy to find that

$$\frac{1}{C_{Vg}} = \frac{1 - DQ}{C_V}. \tag{35}$$

Now that for the pure self-gravitating system, $C_V$ is negative, the TSC of the Tsallis equilibrium becomes

$$Q > \frac{1}{D}. \tag{36}$$

That is to say, when the above inequality (36) is satisfied, the pure self-gravitating system is thermodynamically stable. On the contrary, the system is unstable. According to Eq.(24), the nonextensive parameter $Q$ is determined by the Lagrange temperature gradient and the gravitational potential. So we can find one convenient method to determine whether such a system is stable or not. In the first kind of self-gravitating system, the average value of the parameter $Q$ in the whole system is small enough (in the heterogeneous system, $Q$ in Eq.(34) should be the average value), so the inequality (36) is not satisfied. This implies $C_{Vg}$ and $C_V$ have the same sign. That is why the stability of the first kind of self-gravitating system can also be determined by $C_V$ [15] (keeping in mind that in this situation the sigh of $C_V$ changes due to the phase transition associated with the molecular potentials).

According to Eq.(2), the gravitational thermal capacity is expressed by

$$C_{Vg} = \frac{1}{DQ - 1} \frac{D}{2} N k. \tag{37}$$

This is the gravitational thermal capacity of the pure self-gravitational system (i.e., the second kind of self-gravitational system [15]). From Eq.(37), the sign of $C_{Vg}$ is determined only by the parameter $Q$. Different $Q$ value describes the different velocity $q$-distribution and thus the different Tsallis equilibrium state. However, not all the Tsallis equilibrium states are stable. Only those states satisfying the condition (36) are thermodynamically stable.

What here we should mention is that the evolution exists in the self-gravitating system whose total energy is definite, which leads to the change of $Q$ according to the relationship (24). On the basis of the classical statistics, the isolated self-gravitating system will evolve to the so-called gravothermal catastrophe. Ordinarily, it is thought to be unavoidable now that the thermal capacity is always negative. However, based on Eq.(37) in nonextensive statistics, such a gravothermal catastrophe is avoidable. For the isolated self-gravitating system, after undergoing a series of the Tsallis equilibrium states or the quasi-equilibrium states, it is possible to arrive at one stable equilibrium state. In this process, the isothermal phase transition appears if *DQ* approaches to 1 in Eq. (37). However, the actual evolution process of the self-gravitating system is complicated, so the detailed description for the processes is difficult. This is beyond this paper and remains in the future.

## 6. Conclusions and discussions

In summary, at the beginning of the paper we presented the method in Eq.(7) to estimate the minimum mass $M_L$ of a pure self-gravitating system. For the system consisting of argon gas, the mass $M_L$ is about $1/10^5$ of the Earth's mass. In the system with mass close to the limit mass, the intermolecular potentials can not be neglected, which means the phase transition is unavoidable under certain conditions. In order to avoid the phase transition, the total mass of the self-gravitating system must be large enough. At least, its total mass should reach the order of $10^7 M_L$, which is actually the lowest mass limit of the pure self-gravitating system discussed in this paper.

The pure self-gravitating system, if there is no any heat source around it apart from the vacuum, is thought to be isolated. Here, the concept of 'heat' is related to the molecular motion. The vacuum is not a heat source because it can not exchange heat with the self-gravitating system. It can exchange radiation energy with the system and then induce the self-gravitating system to lose the energy. The energy loss rate is proportional to the Lagrange temperature gradient at the verge of the system. For the second kind of the self-gravitating system, the energy loss rate is very small. The time scale of the energy loss for such a system is very long, so the thermal radiation effect can be neglected ordinarily.

Such a pure self-gravitating system can approach to the hydrostatic equilibrium, satisfying equation (20). The hydrostatic equilibrium is a nonequilibrium stationary state, and under certain condition it is equivalent to the Tsallis equilibrium with the power-law velocity *q*-distribution (22). Based on this equivalence property and the nonextensive parameter equation (24), we introduced a new concept of temperature in Eq.(26) to the self-gravitating systems. It is called the gravitational temperature $T_g$.

The gravitational temperature determines the thermal balance of different parts of the self-gravitating system at Tsallis equilibrium, so it becomes an important state parameter associated with the thermodynamic evolution of the system. And then the concept of the gravitational thermal capacity is also introduced in Eq.(29) as the derivative of the total energy to the gravitational temperature. This new thermal capacity is the vital quantity to determine the evolution of the system. In the pure self-gravitating system, no matter whether it is the first or the second kind, and as long as its gravitational thermal capacity $C_{Vg}$ is positive, the system is thermodynamically stable. The negative thermal capacity drives the system to evolve and eventually induces the system's thermal capacity to be positive.

In the whole paper, we assumed *Q*>0, which is reasonable for the self-gravitating system. According to the relationship (24), *Q*>0 actually reveals the characteristic of the distribution of the Lagrange temperature in the self-gravitating system. In fact, we need to give more restricts to the nonextensive parameter *q*. The parameter *q*>0 is the demand of the entropy increase, and *q*≠1 shows the fact that the velocity distribution function suitable for the self-gravitating system must have the power-law form in Eq.(22). For example, the dark matter density distribution in the galactic halo is one example of the power-law forms [3]. The calculation using the data of standard solar model further restricts the range of $Q$ to be $0<Q<1$ [24].

The condition (25) is very important. Now that the relationship in Eq.(24) always holds as long as the self-gravitating system is at the hydrostatic equilibrium, Eq.(25) gives directly the homogeneous character of the gravitational temperature. That is,

$$\nabla T_g = \frac{m\varphi}{k}\nabla Q. \qquad (38)$$

Obviously, when Eq.(25) is satisfied, the system is at the Tsallis equilibrium. This means that the Tsallis equilibrium is not always identical to the hydrostatic equilibrium. When the system evolves due to thermodynamic instability, it can undergo a series of Tsallis non-equilibrium states, at which Eqs. (21) and (25) don't hold at the same time. In the evolution, the system can be at hydrostatic equilibrium and satisfies the equation (24). Apart from the Tsallis non-equilbirum states, (25) also does not hold in some other situations, such as in the first kind and third kind of self-gravitating systems [15]. Even though, with the generalized local (Tsallis) equilibrium assumption, the gravitational temperature can also be defined. This means we can calculate the gravitational temperature gradient and present a gravitational heat conduction mechanism for the self-gravitating systems [25].

The Lagrange temperature in the relation (28) is a real physical quantity which is measurable in experiments [20], so it is the thermodynamic temperature in the classical BG statistics. Just in this sense, instead of the generalized forms in nonextensive statistics, we adopt the classical forms of the virial theorem and the energy equipartition theorem in Eqs. (32) and (33).

**Acknowledgements**

This work is supported by the National Natural Science Foundation of China under Grant No.11175128 and the Higher School Specialized Research Fund for Doctoral Program under grant No.20110032110058. Zheng Y is also supported by the National Natural Science Foundation of China under Grant No.11405092 and the Heilongliang Province Education Department Science and Technology Research Project No. 12541883.

**References**

[1] C. Tsallis, J. Stat. Phys. **52** (1988) 479.
[2] C.Tsallis, Nonextensive statistical mechanics – an approach to complexity, in G. Contopoulos and P.A. Patsis, *Chaos in Astronomy,* Astrophysics and Space Science Proceedings, P.309-318 and the references therein, Springer-Verlag Berlin Heidelberg, 2009.
[3] M.P. Leubner, Astrophys. J., **632** (2005) L1; M. P. Leubner and Z. Voros, Astrophys. J. **618** (2005) 547.
[4] J. L. Du, Europhys. Lett. **75** (2006) 861; J. L. Du, Astrophys. Space Sci. **312** (2007) 47 and the references therein.
[5] V.F. Cardone, M.P. Leubner, A. Del Popolo, Mon. Not. R. Astron. Soc. **414** (2011) 2265.
[6] M. P. Leubner, Astrophys. J. **604** (2004) 469.
[7] Z. Liu and J. L. Du, Phys. Plasmas **16** (2009) 123707.
[8] M. Tribeche and H. R. Pakzad, Astrophys. Space Sci. **339** (2012) 237.
[9] J.Y. Gong, Z. P. Liu and J. L. Du, Phys. Plasmas **19** (2012) 083706.

[10] J. L. Du, Phys. Plasmas **20** (2013) 092901.
[11] H. N. Yu and J. L. Du, Annals Phys. **350** (2014) 302.
[12] D. Lynden-Bell, Physica A, **263** (1999) 293.
[13] J. Binney and S. Tremaine, Galactic Dynamics, second edition, Princeton University Press, 2011.
[14] A. Taruya and M. Sakagamib, Physica A **307** (2002) 185.
[15] Y. Zheng, EPL **102** (2013) 10009.
[16] Z. C. Wang, Thermodynamics and Statistical Physics, second edition, Higher Education Press, Beijing,1992 (in Chinese).
[17] J. A. S. Lima, R. Silva and A. R. Plastino, Phys. Rev. Lett. **86** (2001) 2938.
[18] J. L. Du, Cent. Eur. J. Phys. **3** (2005) 376.
[19] S. Abe, S. Martínez, F. Pennini and A.Plastino, Phys. Lett. A **281** (2001) 126.
[20] Y. Zheng, Physica A **392** (2013) 2487.
[21] J. L. Du, Europhys. Lett. **67** (2004) 893.
[22] W. C. George, The virial theorem in stellar astrophysics, Pachart Pub. House, 1978.
[23] T. Wada, Phys. Lett. A **297** (2002) 334.
[24] Y. Zheng, EPL **101** (2013) 29002.
[25] Y. Zheng and J. L. Du, EPL **105** (2014) 54002.